\documentclass[aps,prl,twocolumn,superscriptaddress,showpacs]{revtex4-1}
\usepackage{graphicx}
\usepackage{amsmath}
\usepackage{amsfonts}
\usepackage{bm}
\usepackage{color}
\usepackage{ulem}

\newcommand{\ket}[1]{|#1\rangle}

\newcommand{\mrm}[1]{\mathrm{#1}}

\begin{document}
\title{Unusual Two-stage Dynamics of the Spin-Lattice Polaron Formation}

\author{Jan \surname{Kogoj}}
\affiliation{J. Stefan Institute, 1000 Ljubljana, Slovenia}

\author{Zala \surname{Lenar\v{c}i\v{c}}}
\affiliation{J. Stefan Institute, 1000 Ljubljana, Slovenia}

\author{Denis \surname{Gole\v{z}}}
\affiliation{J. Stefan Institute, 1000 Ljubljana, Slovenia}

\author{Marcin \surname{Mierzejewski}}
\affiliation{Institute of Physics, University of Silesia, 40-007 Katowice, Poland}

\author{Peter \surname{Prelov\v{s}ek}}
\affiliation{J. Stefan Institute, 1000 Ljubljana, Slovenia}
\affiliation{Faculty of Mathematics and Physics, University of Ljubljana, 1000
Ljubljana, Slovenia}

\author{Janez \surname{Bon\v{c}a}}
\affiliation{J. Stefan Institute, 1000 Ljubljana, Slovenia}
\affiliation{Faculty of Mathematics and Physics, University of Ljubljana, 1000
Ljubljana, Slovenia}

\begin{abstract}
We follow the formation of a spin-lattice polaron Êafter a quantum quench that simulates
absorption of the pump--pulse in the time--resolved experiments. ÊWe discover a
two-stage relaxation where spin and lattice degrees of freedom represent an integral part of the relaxation mechanism. In the first stage  the kinetic energy of the spin-lattice polaron  relaxes towards its ground state value while relaxation processes via spin and phonon degrees of freedom remain roughly independent. In the second, typically much longer stage, a subsequent energy transfer  between lattice and spin degrees of freedom via the charge carrier emerges. The excess local spin energy radiates away via magnon excitations. 

\end{abstract}
\maketitle

Recent developments in the experimental techniques open unprecedented possibilities
of studying the dynamics of interacting quantum systems. 
The time--resolved spectroscopy of 
solids is one of the most spectacular examples of this progress \cite{matsuda1994,okamoto2010,cortes2011,dalconte12,rettig12,novelli12,kim12}. 
 Solids are complex objects consisting of various subsystems with
different  excitations, like phonon--, spin-- and charge--excitations.
The real--time measurements of the relaxation processes give important 
information about interactions between these subsystems  at various time/energy scales \cite{demsar2003,dexheimer2000,cortes2011,kawakami10}. 
However, the relaxation of various subsystems occurs at exceedingly different  time--scales 
\cite{sensarma2010,zala2012a,eckstein2013,kabanov2008,ku2007,johansson2004,takahashi2002,gumhalter2006}.
In particular, the time--resolved spectroscopy revealed a complex hierarchy of the  relaxation times \cite{dalconte12}  ranging from tens of fs  (e.g., for the coupling between charges and spin excitations)  up to several ps  (e.g., for the coupling between charges and some phonon branches).   Such a broad range of relaxation times poses 
serious challenge for the theoretical analysis: {\it (i)} due to a large number of relevant degrees of freedom studying several subsystems is a complicated task itself  {\it (ii)} some theoretical approaches (e.g. $t$-DMRG) are useful/applicable only in a certain time--window. 
Hence, various stages of the relaxations have been investigated within very different (and mostly non--overlapping) approaches \cite{white2004,jim2006,fehske09,schiro10,Manmana2007,Knap2013,defilippis12}. The initial ultrafast  stage has been studied within fully quantum and fully nonequilibrium approaches \cite{takahashi2002,matsueda12,golez2013}. Most of the up--to-date approaches  take into account  charge carriers that  couple to only one type of chargeless excitations. The subsequent slower stage consist in the energy flow between various chargeless subsystems and can be studied within  quasiequilibrium approaches which rely on the notion of well defined temperatures \cite{dalconte12}. 

In this {\it Letter} we apply a single fully nonequilibrium approach 
to show  how the multi--stage relaxation emerges in systems of a charge carrier coupled to magnons and phonons. The initial ultrafast cooling of highly--excited charge is followed by much slower exchange of energies   between the  magnon  and the phonon subsystem. Despite the absence of any direct coupling  between   magnons and phonons  the latter stage of relaxation can be effectively mediated even by very dilute charge carriers.   

We consider a single  hole within the $t$-$J$ Holstein model in one spacial dimension under the influence of a staggered field:
\begin{eqnarray}
 H&=&H_\mathrm{kin}+H_J  +H_h +H_{\mathrm EP} +H_{\mathrm ph},  \nonumber \\
  H &=& -t_{0}\sum_{i,\sigma} [\tilde c_{i,\sigma}^{\dagger} \tilde c_{i+1,\sigma} + \mbox{H.c.} ] + J\sum_{i}\mathbf{S}_{ i} \cdot \mathbf{S}_{ i+1}, \\
    &+& h\sum_i (-1)^iS_i^z + {g} \sum_{{i}} n_{i}^\mrm{h} (a_{{i}}^\dagger + a_{{i}}) 
  +  \omega_0\sum_{{i}} a_{{i}}^\dagger  a_{{i}},\nonumber
\label{ham} 
\end{eqnarray}
where $t_{0}$ is the nearest neighbor hopping amplitude,  $\tilde c_{i,\sigma}=c_{i,\sigma}(1-c_{i,-\sigma}^{\dagger}c_{i,-\sigma})$ is a projected fermion operator, $J$ represents the Heisenberg exchange interaction,   $\mathbf{S}_{i}$ is the spin operator and $h$ represents the staggered magnetic field. Electron phonon  coupling strength is given  by $g$,    $a_{i}^{\dagger}(a_i)$ are phonon creation (destruction) operators at sites $i$, and $n_{i}^\mrm{h}=1-\sum_\sigma\tilde c_{i,\sigma}^{\dagger}\tilde c_{i,\sigma}$ is the hole density. $\omega_{0}$ denotes the dispersionless phonon frequency. We measure all quantities in units of $t_0$ and finally set $t_0=1$. The main reason for including the staggered field is to remove the spin-charge separation thus introducing the notion of the string picture, characteristic of the two-dimensional system.

We  employ  the  exact diagonalization method (ED) defined over a limited functional space (EDLFS), which was successfully used to describe properties of a carrier doped into a planar ordered AFM described by the $t$-$J$ model \cite{bonca2007} and in the presence of lattice degrees of freedom, \cite{lev2011,lev2011_1}. 
The advantage of EDLFS over the standard ED follows from systematic generation of states which contain spin and phonon excitations in the vicinity of the carrier.
We  compute  the initial  state $\ket{\psi(t=0)}$ using the Lanczos technique and setting the initial value of the overlap integral to $t_0=0$.
We then make a sudden quench by switching $t_0$ from 0 to 1  and time evolve the initial  $\ket{\psi(0)}$   using the  time propagator with  the quenched Hamiltonian. At each small time step $\delta t\ll 1$ we use  Lanczos basis for generating the evolution $\ket{\psi(t-\delta t)}\rightarrow\ket{\psi(t)}$ \cite{mierzejewski2010,mierzejewski2011,park1986}.

\begin{figure}
\includegraphics[width=0.43\textwidth]{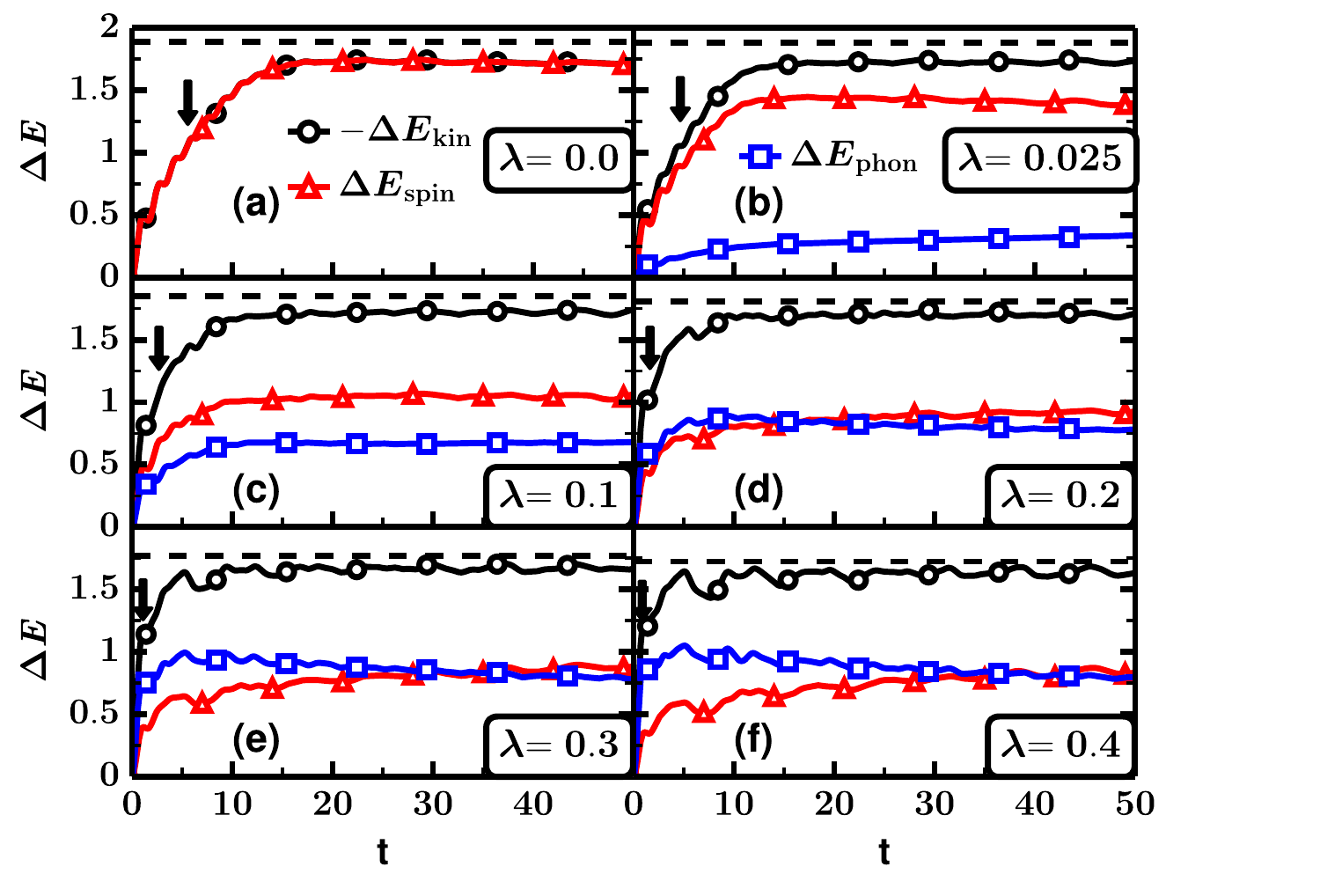}
\caption{The evolution after the $t_0$ quench: time--dependence of the change of the kinetic energy of the hole (circles)  $\Delta E_\mrm{kin}=\langle\psi(t)\vert H_\mrm{kin}\vert\psi(t)\rangle-\langle\psi(0)\vert H_\mrm{kin}\vert\psi(0)\rangle$ (note that we plot $-\Delta E_\mrm{kin}$), total spin energy (triangles)  $\Delta E_\mrm{spin}=\langle\psi(t)\vert H_{J}+H_{h}\vert\psi(t)\rangle-\langle\psi(0)\vert H_{J}+H_{h}\vert\psi(0)\rangle$, phonon energy (squares) $\Delta E_\mrm{phon}=\langle\psi(t)\vert H_\mrm{ph}+H_\mrm{EP}\vert\psi(t)\rangle-\langle\psi(0)\vert H_\mrm{ph}+H_\mrm{EP}\vert\psi(0)\rangle$ for different values of $\lambda=g^2/2\omega_0t_0$,  $J=0.3$,  $h=0.7J$ and $\omega_0=1.0$. Horizontal dashed lines represent absolute values of kinetic energies, calculated in the respective ground states of the quenched Hamiltonians (with $t_0=1$).  Vertical arrows indicate $\tau_\mrm{form}$. 
}
\label{fig1}
\end{figure}

In Fig.~\ref{fig1} we display different energies representing expectation values of different parts of Hamiltonian in Eq.~\ref{ham} during the time evolution.  Immediately after the quench the kinetic energy is zero, {\it i.e.}  $E_{\mathrm{kin}}=\langle\psi(0)\vert H_{\mathrm {kin}}\vert\psi(0)\rangle = 0$, since the initial state was prepared using $t_0=0$. During the time evolution the kinetic energy decreases and approaches its respective value of the spin-lattice polaron (SLP)  in the  ground state at $t_0=1$.  Since the total energy remains a constant after the quench,  the sum of all other energies must increase. 
Although our system is not coupled to an external electromagnetic field, its relaxation after the $t_0$ quench
is similar to the relaxation after the absorption of the electromagnetic pulse.
In both cases the essential physics consists in transforming the highly elevated kinetic energy into other excitations.
This claim is supported by explicit numerical simulations in Ref. \cite{golez2013}.
In the $\lambda=0$ case, the decrease of $\Delta E_{\mathrm{kin}}$  is exactly mirrored by the increase of the total spin energy, $\Delta E_\mathrm{spin}=\Delta E_J + \Delta E_h$, see Fig.~\ref{fig1}(a). During the spin polaron formation, manifested in the decrease of $\Delta E_\mrm{kin}$, excess energy is absorbed by the inelastic spin degrees of freedom, directly coupled to the hole. We refer to the time scale at which the kinetic energy  saturates as the spin polaron formation time  $ \tau_\mrm{form}$. 
It is formally obtained from fitting the kinetic energy to a functional form $ \Delta E_\mrm{kin}(t)=a[1-\exp(-t/\tau_\mrm{form})]$. The values of $\tau_\mrm{form}$ are indicated by vertical arrows in Fig.~\ref{fig1}.

Switching on  EP  coupling  adds  additional degrees of freedom, coupled to the hole. The most discernible effect of increasing $\lambda$ is the shortening of the SLP formation time $ \tau_\mrm{form}$, qualitatively consistent with the Matthiessen's rule. For example, at  small $\lambda=g^2/2\omega_0=0.025$,  $\Delta E_{\mathrm{kin}}$  decreases slightly faster in comparison to the  $\lambda=0$ case. Moreover, the excess energy  is  distributed between $\Delta E_\mathrm{spin}$ and  $\Delta E_\mathrm{phon}=\Delta E_\mrm{ph}+\Delta E_\mrm{EP}$, as seen in  Fig.~\ref{fig1}(b). 
Following more closely the time evolution of $\Delta E_{\mathrm{spin}}$ we observe that  $\Delta E_{\mathrm{spin}}$ reaches a broad maximum just above  $t\gtrsim  \tau_\mrm{form}$, which is followed by  a gradual  decrease that  is matched by a slow increase in  $\Delta E_\mathrm{phon}$. During this time $\Delta E_{\mathrm{kin}}$ remains largely  unchanged. These results are consistent with  a subsequent slow redistribution of the energy  from  spin  to lattice degrees of freedom. While there is no direct coupling between the spin and the lattice sector,  such redistribution can only take place via coupling to the charge.  It  is thus not surprising that we obtain a  much longer time scale for this energy exchange process, with a very rough estimate  $t_\mrm{ex}\gtrsim 50$.

With further increasing  $\lambda$, the SLP formation time  $ \tau_\mrm{form}$ further shortens and up to $\lambda=0.1$ the final amount of the excess energy absorbed by phonons, $\Delta E_\mathrm{phon}$,  increases. In this particular  case, see Fig.~\ref{fig1}(c), we observe no change of different parts of energies after $ \tau_\mrm{form}$.  A different physical picture is seen in the case when $\lambda\gtrsim 0.2$. In this case the $\Delta E_\mathrm{kin}$ again reaches  the steady state value within the initial time $t\sim \tau_\mrm{form}$, meanwhile $\Delta E_\mathrm{phon}$ reaches a broad maximum. However, with further increasing of time,  $t\gtrsim \tau_\mrm{form}$, we observe a subsequent energy flow, which is in this case reversed in comparison to $\lambda=0.025$ case, {\it i.e.} from (decreasing) $\Delta E_\mathrm{phon}$ to (increasing) $\Delta E_\mathrm{spin}$. This energy transfer from lattice to spin degrees o freedom again takes place on a much longer time scale $t_\mrm{ex}$ in comparison with the  relaxation time of the kinetic energy, {\it i.e.} $t_\mathrm{ex}>>\tau_\mrm{form}$. The amount of the subsequent energy transfer    becomes more pronounced at larger $\lambda=0.4$. Again,  during this  energy transfer the kinetic energy of the SLP remains roughly unchanged. Moreover, comparing $\Delta E_\mathrm{phon}(t)$  in the long-time limit, $t\sim 50$,   for systems with increasing $\lambda\gtrsim 0.1$, we find that  the   energy  absorbed by the lattice  saturates as $\lambda$ increases  towards $\lambda=0.4$.  



\begin{figure}
\includegraphics[width=0.43\textwidth]{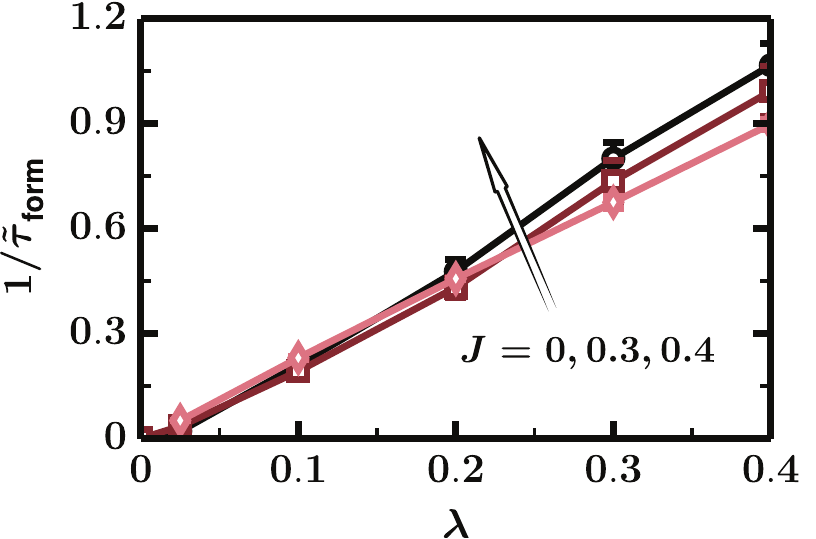}
\caption{${1/\tilde \tau_\mrm{form}}(\lambda)$ obtained from Eq.~\ref{matt} using  $J=0,0.3$ and 0.4 with $h=0.7J$.   In the case of $J=0$ we obtain  $\tau^{-1}_\mrm{form}(J=0,\lambda=0)=0$ since in this case  the hole behaves as a free particle which leads to an infinite relaxation time. Results for $J=0$ were then obtained from $\tilde \tau_\mrm{form}(\lambda)=\tau_\mrm{form}(J=0,\lambda)$.
}
\label{fig2}
\end{figure}

While the  subsequent energy transfer between spin and lattice degrees of freedom clearly indicates that the emission/absorbtion  of phonons and spin excitations represents strongly interconnected processes  in the second stage of relaxation, there remains an open question concerning the interdependence of these inelastic processes in the first stage of the relaxation.  
To gain further insight  into  scattering process in the first stage of relaxation we
test applicability of  the Matthiessen's rule and split the SLP formation time  into two possibly independent contributions:
\begin{equation} 
{\tau^{-1}_\mrm{form}}(J,\lambda) = {\tau^{-1}_\mrm{form}}(J,\lambda=0)+ {\tilde \tau^{-1}_\mrm{form}}(\lambda).
\label{matt}
\end{equation}
Based on the assumption of the validity of the Matthiessen's rule $\tilde\tau_\mrm{form}(\lambda)$ represents the bare phonon contribution to the SLP formation time.  In the case of independent scattering processes  $\tilde\tau_\mrm{form}(\lambda)$ should remain $J-$independent. In Fig.~\ref{fig2} we present $\tilde \tau^{-1}_\mrm{form}(\lambda)$ extracted from systems with  different values of $J$.  Up to $\lambda\lesssim 0.2$ the values nearly overlap, signaling that the emission of phonons and local string excitations represent nearly independent processes. For larger $\lambda=0.3$ and 0.4, we observe a slight upward deviation of $1/\tilde \tau_\mrm{form}(\lambda)$ for systems with increasing  $J$. 

We shall gain additional  insight into this unusual relaxation dynamics by computing time-dependent change of the hole-spin and hole-phonon number correlation functions defined as:

\begin{multline}
C_s(t,j)=\sum_{i}(-1)^{i+j}\Bigl[ \langle\psi(t)\vert n_i^\mrm{h}S_{i+j}^z\vert\psi(t)\rangle \\ 
            - \langle\psi_\mrm{G}\vert n_i^\mrm{h}S_{i+j}^z\vert\psi_\mrm{G}\rangle\Bigr]\label{cs}
\end{multline}

\begin{multline}
C_\mrm{ph}(t,j)=\sum_{i}\Bigl[ \langle\psi(t)\vert n_i^\mrm{h}n_{i+j}^\mrm{ph}\vert\psi(t)\rangle \\
            -  \langle\psi_\mrm{G}\vert n_i^\mrm{h}n_{i+j}^\mrm{ph}\vert\psi_\mrm{G}\rangle\Bigr]\label{cph},
\end{multline}
where $n_{i}^\mrm{ph}=a_i^+a_i$ is the phonon number operator and $\vert\psi_\mrm{G}\rangle$ is the ground state wavefunction of the quenched Hamiltonian. 

\begin{figure}
\includegraphics[width=0.43\textwidth]{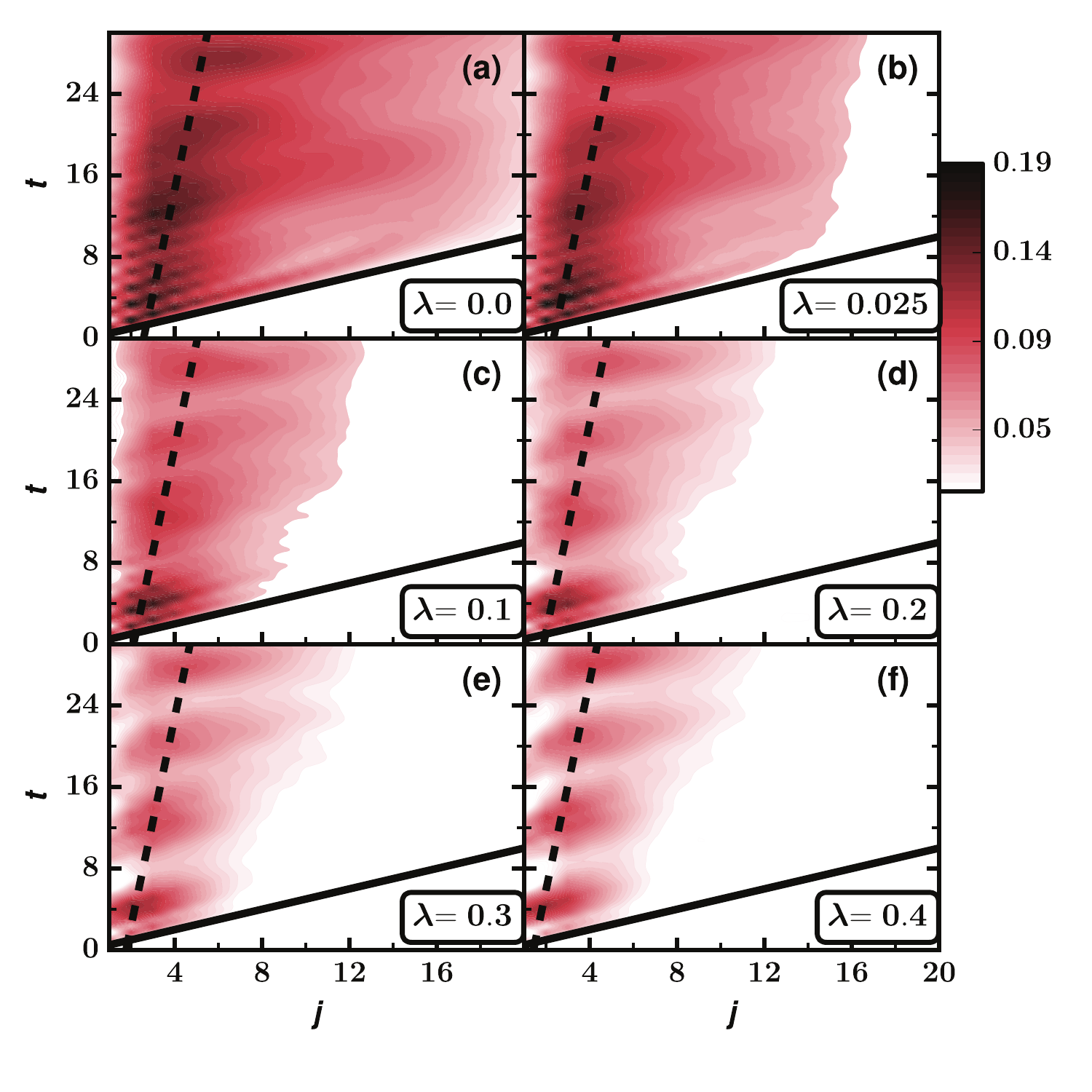}
\caption{Hole-spin correlation function $C_s(t,j)$ for $J=0.3$, $h=0.7J$ and different values of $\lambda$. Full line indicates maximal free electron velocity $v_\mrm{free}=2$ and the dashed line represents the magnon velocity $v_\mrm{mag}=J+h-\sqrt{h(h+2J)}$.
}
\label{fig3}
\end{figure}

{ In Fig.~\ref{fig3}(a) we present the density plot of the hole-spin correlation function representing the distribution of spin excitations relative to the hole position at $j=0$ for the case of $\lambda = 0.0$. At short times,  $t\lesssim \tau_\mrm{form}$, we observe a fast expansion of the front line of $C_s(t,j)$  with a well defined initial velocity approximately equal  the maximal group velocity of the free particle, $v_\mrm{free}\sim 2$, indicated by a full straight line. The expansion at later times $t\gtrsim \tau_\mrm{form}$ slows down. The peak  values of $C_s(t,j)$ separate from the hole position at $j=0$ and move away with a rather well defined velocity   that approximately matches  the maximal magnon velocity $v_\mrm{mag}=J+h-\sqrt{h(h+2J)}$, as indicated by the dashed line.  } 

{ The emerging physical picture is consistent with  a  two-stage spin polaron formation process. In the first stage  the hole travels   with the velocity not exceeding  the free particle one,  $v_\mrm{free}\sim 2$, and emits  its excess  energy by creating local spin excitations.  This stage is completed  in a very short time,  roughly given by  $\tau_\mrm{form}$,  as the kinetic energy of the hole approaches the kinetic energy of the spin polaron in its ground state.  An excited spin polaron is thus formed. At this point the polaron  is surrounded by the excess amount of local spin excitations. In the second stage the dissipation emerges  through a process where  the  excess spin energy is   radiated away via magnon excitations while $\Delta E_\mathrm{kin}$ remains nearly constant.
}

{
This effect survives as well at finite values of $\lambda$, see Figs.~\ref{fig3}(b-f). In this case,  as already seen from  Fig.~\ref{fig1}, a part of the excess kinetic energy is absorbed by the phonon subsystem, which renders less available energy for spin excitations. The most discernible effect of increasing $\lambda$ on the spin subsystem  is thus the overall decrease of $C_s(t,j)$. Due to the absence of direct magnon-phonon coupling,  $v_\mrm{mag}$  remains unchanged. The other noticeable effect of increasing $\lambda$ is the shortening time of the initial fast expansion of $C_s(t,j)$ with $v_\mrm{free}$. 
}

\begin{figure}
\includegraphics[width=0.43\textwidth]{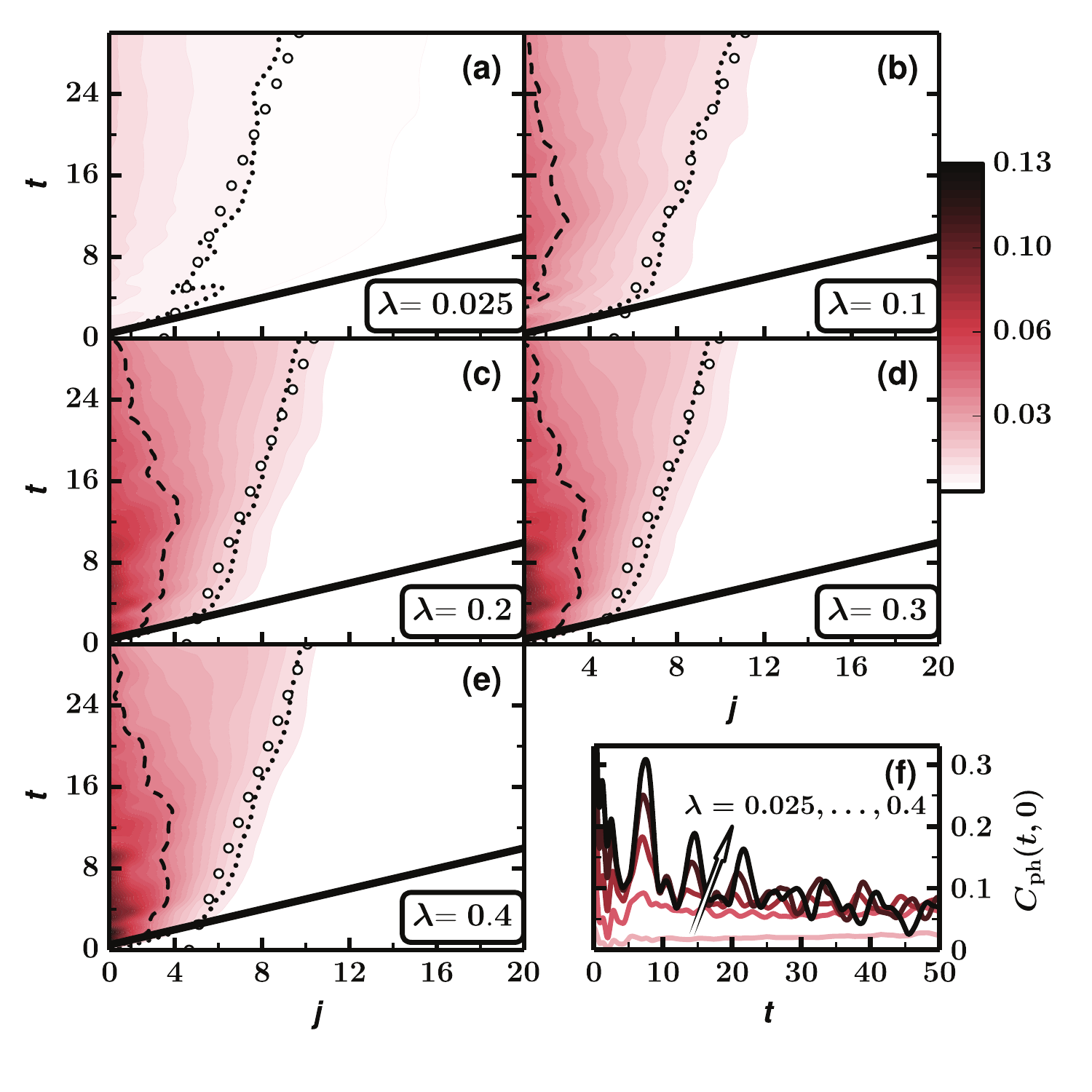}
\caption{Hole-phonon number correlation function $C_\mrm{ph}(t,j)$ (density plots (a-e) and on-site value (f)) the rest of parameters are the same as in Fig.~\ref{fig3}. Dashed and dotted lines connect points  of selected constant  values of $C_\mrm{ph}(t,j)$. Full line indicates maximal free electron velocity $v_\mrm{free}=2$ and open circles present $j_\mrm{ave}(t)$ as explained in the text. 
}
\label{fig4}
\end{figure}

In Fig.~\ref{fig4} we show the density plot of the hole-phonon number correlation function.   As in the previous case, we observe a fast expansion of $C_\mrm{ph}(t,j)$ for short times, $t\lesssim \tau_\mrm{form}$ with a well defined velocity given by $v_\mrm{free}$. At longer times, $t\gtrsim \tau_\mrm{form}$, two distinct effects are observed: {\it (i)} a decrease of $C_\mrm{ph}(t,j)$ at smaller distances from the hole position, $j\lesssim 6$,  as highlighted  by a dashed lines in Fig.~\ref{fig4} and {\it (ii)} a further  expansion of $C_\mrm{ph}(t,j)$ with rather well defined velocity at larger distances  for $j\gtrsim 6$, emphasized  by dotted lines. Since  Einstein phonons possess zero group velocity, there must exist  an alternative mechanism for the explanation of the observed  velocity. Within a semi-classical picture   the highly excited hole first  slows  down by creating local spin and phonon excitations thus forming an excited SLP with its  kinetic energy $ E_\mrm{kin}(t)$ that remains  above its equilibrium  ground state  value, $E_\mrm{kin}(t)=\Delta_\mrm{kin}(t)+E_{\mrm{kin},G}$, see also Fig.~\ref{fig1}. The existence of the finite  $\Delta_\mrm{kin}(t)$ is not solely due to finite--size effects but primarily due to a finite energy gap in the system.  For this reason the relaxing SLP does not approach  its exact ground state since it can not emmit arbitrarily small energy quantum. Due to the finite value of  $\Delta_\mrm{kin}(t)$, the average group velocity of the excited SLP $v_\mrm{ave}$ remains finite. The excited SLP in the semi-classical sense  moves away from its otherwise localized phonon excitations, which in turn causes the observed expasion of $C_\mrm{ph}(t,j)$. To test this idea   we estimated  the averaged SLP group velocity from:
\begin{equation}
v_\mrm{ave}= \sqrt{{1\over \pi}\int_{-\pi/2}^{\pi/2} \mrm{d}k\left ( {\mrm{d}E(k)\over \mrm{d}k}\right )^2},
\end{equation}
where $E(k)$ is the equilibrium SLP dispersion relation  computed using the quenched Hamiltonian while the average is taken over the whole AFM Brillouin zone.   Note that the ground state of the SLP has $k=\pi/2$. Circles in Fig.~\ref{fig4} represent shifted distances $j_\mrm{ave}(t)=tv_\mrm{ave}+ j_0$.  With increasing $\lambda$,   $v_\mrm{ave}$ decreases since the  quasiparticle band becomes narrower. All this is well reflected  by   $j_\mrm{ave}(t)$ in Fig.~\ref{fig4}(a) through (e),  matching rather well the expansion of $C_\mrm{ph}(t,j)$.


The subsequent decrease of the number of phonons observed in  $C_\mrm{ph}(t,j)$ in the vicinity of the hole position, {\it i.e.} for $j\lesssim 6$ for  $\lambda \geq 0.1$, reveals an intricate thermalization process in which the  excited lattice subsystem transfers some of its excess energy to the excited SLP,  which consequently thermalizes  via  magnon emisson. The effect becomes more pronounced with increasing $\lambda$. Taking into consideration the  sum-rule $\sum_jC_\mrm{ph}(t,j)=\Delta E_\mrm{ph}(t)/\omega_0$, we find this behavior consistent with the decrease of $\Delta E_\mrm{ph}(t)$ and its transfer to $\Delta E_\mrm{spin}(t)$, as seen  in Fig.~\ref{fig1}.  The $C_\mrm{ph}(t,0)$, presented in Fig.~\ref{fig4}(f), in the regime of $\lambda\geq 0.2$ clearly shows a decrease of the average on-site phonon number towards the respective  ground state value on the  time-scale $t_\mrm{ex}$. The thermalization of the excess of local phonon  excitations and the energy transfer between the phonon and the spin sub-system  occurs on the same time-scale.

In summary,  the SLP relaxes in two stages. In the first stage  the highly exited hole lowers its kinetic energy by emitting  local spin and phonon excitations and forms an excited SLP  with a non-zero average group velocity and with the  kinetic energy close to its equilibrium value. 
There is no essential departure from the Mattiessen's rule in this stage hence phonons and spin excitations act as  rather independent relaxation mechanisms for highly excited charges. Our results indicate that the  experimentally observed times of the primary relaxation ($\sim 10 fs $) may be explained either within purely magnetic interactions or in a scenario where both phonons and spins couple to charge carriers.
In the second stage of relaxation a subsequent energy transfer between phonon and spin degrees of freedom, mediated by the excited SLP, takes place. Simultaneously, SLP thermalizes by emitting magnons. Thermalization of locally excited phonons is realized in part via the energy transfer from phonons to magnons mediated by the SLP as well as by the spread of the excited phonon cloud that occurs due to a non-zero average group velocity of the SLP.  The second   stage  relaxation time may be  much  longer  than the first stage  one.  
 
 
\begin{acknowledgments}
J.B. acknowledges stimulating discussions with S.A. Trugman and the  support by the P1-0044 of ARRS, Slovenia. M. M. acknowledges support from the NCN project DEC-2013/09/B/ST3/01659.  This work was performed, in part, at the Center for Integrated Nanotechnologies, a U.S. Department of Energy, Office of Basic Energy Sciences user facility. 
\end{acknowledgments}

\bibliography{Polarons.bib}
\end{document}